# Understand Need "Uncertainty Analysis" in the System Development Modeling Process

Kardile Vilas Vasantrao

Department of Computer Science.
Tulajaram Chaturchand College, Baramati
Affiliated by Pune University, Pune, Maharashtra ( India)
(vilaskardile@gmail.com)

**Abstract:** *Software project development process is requiring accurate software cost and schedule estimation for achieve goal or success. A lot it referred to as the "Intricate brainteaser" because of its conscience attribute which is impact by complexity and uncertainty, Generally estimation is not as difficult or puzzling as people think. In fact, generating accurate estimates is straightforward—once you understand the intensity of uncertainty and module which contribute itself process. Because Design and chose approach is repeated incident in our daily life when we plan to go to our work .We estimate the time and risk need for design approach. The estimated time and risk fluctuates according external uncertain factor and theme's condition. In our everyday life, we enhance our estimation based on past experience in which problem solve by which method and in which condition and which opportune provide that method to produce better result .*

*So, Instead of unexplained treatises and inflexible modeling techniques, this will guide highlights a proven set of procedures, understandable formulas, and heuristics that individuals and complete team can apply to their projects to help achieve estimation ability with choose appropriate development approaches*

*In the early stage of software life cycle project manager are inefficient to estimate the effort, schedule, cost estimation and its development approach .This in turn, confuses the manager to bid effectively on software project and choose incorrect development approach. That will directly effect on productivity cycle and increase level of uncertainty. This becomes a strong cause of project failure. So to avoid such problem if we know level and sources of uncertainty in model design, It will directive the developer to design accurate software cost and schedule estimation. which are require l for software project success.*

*This paper demonstrates need of uncertainty analysis module at the modeling process for assist to recognize modular uncertainty system development process and the role of uncertainty at different stages in the modeling processes.*

## Introduction:

Design and chose approach is repeated incident in our daily life when we plan to go to our work .We estimate the time and risk need for design approach. The estimated time and risk fluctuates according external uncertain factor and theme's condition.

In our everyday life, we enhance our estimation based on past experience and historical data.

Design of software design approach is crucial because of today's dynamic environment of software development firm. The World Wide Web has provided a dais for companies to communicate and transact directly with their customers and partners. Challenges arise due to fast evolving technology and increased competition as companies are under constant pressure to develop new functionalities to satisfy changing client needs and to deliver them in short cycles at low costs. (Iansiti and MacCormack 1997).

Existing methods for software development have lot of options which can be classified into two categories, plan-driven (traditional) and practice –driven (Boehm and Turner 2003; Iansiti and MacCormack 1997). At the early stage of the plan-driven approach, the user needs are identified; requirements for new functionalities are specified; technical specifications are created; development processes are defined; specific project targets are spelled out; and, acceptance criteria and tests are outlined. Many of the CMM or ISO based methods belong to this category. The focus of the project team is on development and implementation according to the plan. As a result, the success of a project using plan-driven approach hinges on the validity and reliability of the project plan. On other hand if requirement of user in changeable form then it will increases level of uncertainty the plan driven approach is less effective. The practice–driven (Agile) methodology is characterized by a chaotic perspective, collaborative values and principles, and barely sufficient technology (Highsmith 2002). Its foundation values include: individuals and interactions over processes and tools; working software over comprehensive documentation; customer collaboration over contract negotiation; responding to change over following a plan (Manifesto for Agile Software Development). Compared with the plan-driven approach, the agile methodology addresses the lack of knowledge of both the client (on the technology and development process) and the developer (on client's business needs) by encouraging closer collaboration. Usually, client representatives are collocated and work alongside the project team. Instead of working against the plan, frequent changes are embraced to address the client's changing business needs. As a result the success of project using practice–driven approach hinges on the effective communication skill and correct abstraction formalization of development Team.

In such situation development practitioner and Organization and company has confuse for which approach is abandoning or which adopting because of the strength and weaknesses which will force to learner for accept *"Technology never fail it will fail to produce best result due to opponent opportune."*

**So if developer get to know at early stage of estimation that which methodology is suited for which module of system then we are able to reduce failure cause of software process.**

The aim of this paper is to understand modeling process and uncertainty appearance in particular module for better to estimation process. In section II introduce various factors which affect software productivity. In section III Available software development approaches comparatives. In section IV Modeling process

## Section –II:

Problem: Design a Process**:** In software development, Modern "lightweight" methodologies are gaining ground on more traditional "heavyweight" methodologies. Both have their advantages and disadvantages, and appropriateness where we get best result. Many project fail because of inaccurate handling design approach or decision made early some time prove to be wrong later on. The most critical and crucial part of software development approach is when planning of design development is required in the early stage of the software life cycle where problem to be solved. Estimated, requirement by user is not completely understand and problem to be solved had not yet been completely revealed. The Major issue that separates the various processes that we looked at is the amount of up-front Planning they require. We can think of this as a spectrum, which at one end has a purely Plan oriented and other end practice oriented question is then for any given situation how do find the right approach.

Alternatively You can find a situation where the approach will give best result for this causes to handle this uncertainty we must be understand Risk handle strategies of each approach. Software Risk although there has been considerable debate about proper definition for ware risk, There is general agreement that risk always involves two characterizes:

Uncertainty and Complexity (Project Risk, Technical Risk, Business risk, .etc). Which will directly affect the Software deployment Process and approach? There are four broad control factors. This factor s is interconnected .when one changes at least one other factor must also change.

- Cost- or Effort. Available money impact the amount of effort put into the system
- Schedule – A Software project is impacted as the timeline is changed.
- Requirements-The scope of the work that needs to be done can be increased or decreased to affect the project.
- Quality – Cut control by reducing quality.

To avoid such problem if we know level and sources of uncertainty in model design, initial phase of development , It will directive the developer to design accurate software cost and schedule estimation. Which are essential for software project success . However once the required efforts have estimated, little is done to recalibrate and reduce the uncertainty of the initial estimates.

## Section –III:

*This comparison Focus on :* Practice driven is sometimes characterized as being at the opposite end of the spectrum from "plan-driven" or "disciplined" methods. This distinction is misleading, as it implies that agile methods are "unplanned" or "undisciplined". A more accurate distinction is that methods exist on a continuum from "adaptive" to "predictive". Practice-driven lie on the "adaptive" side of this continuum.

Adaptive methods focus on adapting quickly to changing realities. When the needs of a project change, an adaptive team changes as well. An adaptive team will have difficulty describing exactly what will happen in the future. The further away a date is, the vaguer an adaptive method will be about what will happen on that date. An adaptive team can report exactly what tasks are being done next week, but only which features are planned for next month. When asked about a release six months from now, an adaptive team may only be able to report the mission statement for the release, or a statement of expected value vs. cost.

Predictive methods, in contrast, focus on planning the future in detail. A predictive team can report exactly what features and tasks are planned for the entire length of the development process. Predictive teams have difficulty changing direction. The plan is typically optimized for the original destination and changing direction can cause completed work to be thrown away and done over differently. Predictive teams will often institute a change control board to ensure that only the most valuable changes are considered.

## Categories wise best practice

| Plan Driven Software Development | Practice –driven Software Development |
|---|---|
| High Criticality | Low Criticality |
| Junior Developers | Senior Developer |
| Requirements do not change Often | Requirement change often |
| Large number of Development | Small number of developer |
| Culture that demands order | Culture that thrives on chaos |

**Section –IV: Modeling as a part of Project planning in system development** is one of the most critical activities within the project lifecycle. **Project plan development** is the main part of Project Planning Stage. The project manager takes the responsibility for creating a project plan that is a formal document showing the basis upon which to assess the performance of the project and measure its results. Let's review the steps of **project plan development** in detail.

Create the Work Breakdown Structure.

To design project plan, developer will need to settle on the *Work Breakdown Structure* (the acronym "WBS") for his/her project. Which is give detailed list of all the phases, activities and jobs required for successful project completion .In other word we can the WBS becomes the foundation for his/her project plan as S/he can use it to decide the resources required to deliver each activity or task listed. The WBS allows designer to design simple to-do lists and task lists and then assign them to members of the project team.

The benefit of WBS:

When designer try to developing a project plan WBS depicts the dependencies between tasks. Which is focus on, how each task is associated with other tasks and what (internal or external) dependencies can be set.

1. Define the Required Resources.

Once the plan and its activities required for deliver to project are situate, next step in creating a project plan is to identify the resources required for each task and activity. WBS showing the possibility of the project which describes resources are required and in which quantities and measures. In this step the project resource base and types of resources will be describe.

Generally project may require following resources:

- Full-time and part-time employees
- Equipment and materials

Designer of plan is calculate how many people S/he should employ to do project and Human and equipment resource.

2. Design a Project Schedule.

After the Work Breakdown Structure is completely design. Designer are able create a project plan in which schedule tasks and deadline for activities listed in the

To build a project schedule the following information is necessary:

- Identified tasks and activities and their dependencies (both internal and external)
- Assignments to members of the project team (who will do which task)
- Risk mitigation strategies and a contingency plan  Critical milestones
- Allocated resources required for the project

Designer can design a project schedule on the basis on that information which is described at last two step of project plan development.

But if designer have knowledge of modular uncertainty in work break structure he is able to choose appropriate team for suitable module. But unfortunately there is no such phase who will assist to designer to take proper decision while make project plan

At two previous **project plan development** steps this information has been identified so you can design a project schedule.

A typical modeling study will involve the following four different types of actors:

*Organization environment*
*System Designer and Analyst*
*System Testing*
*End User or stakeholder*:

The modeling process may be different according to the organization.

Fig 1.

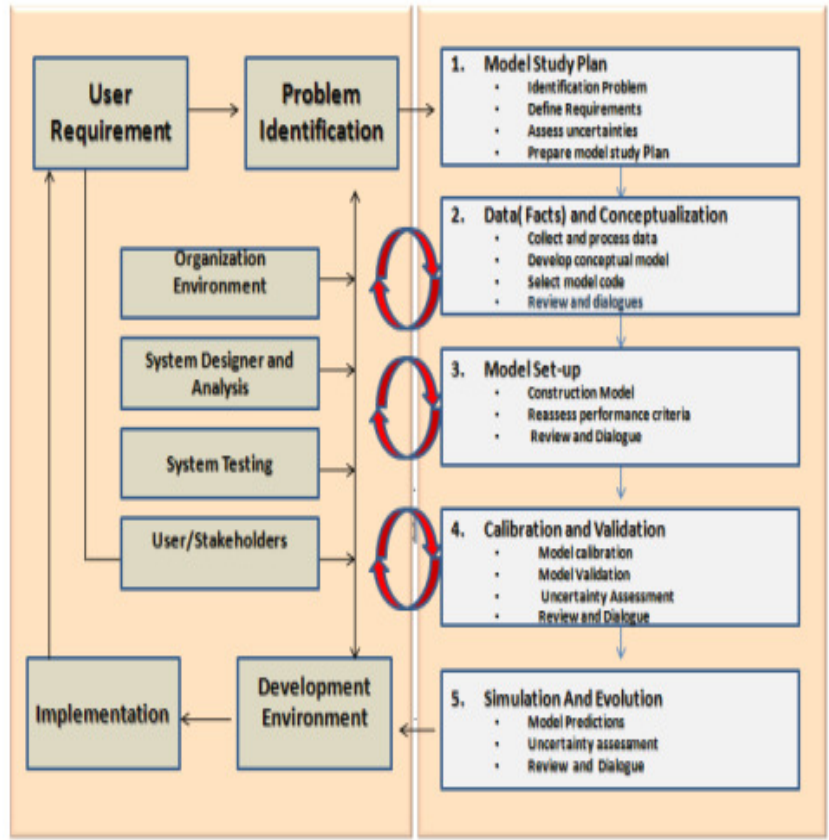

**Step 1:** *Model study plan*            **Step 4:** *Calibration and Validation*

**Step 2:** *Data and conceptualization*  **Step 5:** *Simulation and evaluation*

**Step 3:** *Model set-up*

**Section –IV:**

4 .1 Uncertainty expressions and classification

4.1. A Definitions
Uncertainty and associated terms such as error, risk and ignorance are defined and interpreted differently by different authors; see Walker et al. (2003) for a review. There are different definitions available in various literature.

we adopt a subjective understanding of uncertainty in which the degree of confidence Thus according to our definition a person is uncertain if s/he lacks confidence about the specific outcomes of an event. Reasons for this lack of confidence might include decision of the information as incomplete, unclear, inaccurate, unreliable, inconclusive, or potentially false. Similarly, a person is certain if s/he is confident about the outcome of an event. It is possible that a person feels certain but has take wrong decision of the information

There are many different decision situations, with different possibilities for characterizing uncertainty. Uncertainty is also known degree of unreliability of knowledge, which translates into a state of confidence.

**4.1.B Classification**:

Classification of unsatisfactory knowledge is represented by Brown (2004) . Which is useful to differentiate between bounded uncertainty and unbounded uncertainty

<u>Bounded uncertainty:</u> In which all possible outcomes are deemed 'known' and we can define only quantitative probabilities require all possible outcomes of an uncertain event and each of their individual probabilities to be known.

<u>Unbounded uncertainty:</u> In which some or all possible outcomes are "deemed unknown".

**4.2. Sources of uncertainty**

Uncertainty is noticeable itself at different locations in the model-based software project management process.
In model base project management process uncertainty is noticeable at different location.
This can characterize as : Context and framing,

**Context:** At the initial stage of problem phase where problem is understand the model context is defined. Which include external circumstance like the technological external economic, environmental, political, social circumstances that form the context of the problem.
Input uncertainty in terms of external driving forces (within or outside the control of the software project manager) and system data that drive the model.

International Journal of Software Engineering & Applications (IJSEA), Vol.2, No.3, July 2011

**Framing:** Model structure uncertainty is conceptual uncertainty is because of partial understanding and easy descriptions of modeled processes as compared to reality.

Parameter uncertainty is the uncertainties related to parameter values.

Model technical uncertainty is the uncertainty arising from computer implementation of the model, because of numerical approximations, resolution in space and time, and bugs in the software.

From the all above sources the total uncertainty on the model simulations and model output uncertainty, can be assessed by uncertainty propagation

Fig. 2 Different uncertainty situation and categorization of imperfect knowledge resulting (Brown, 2004).

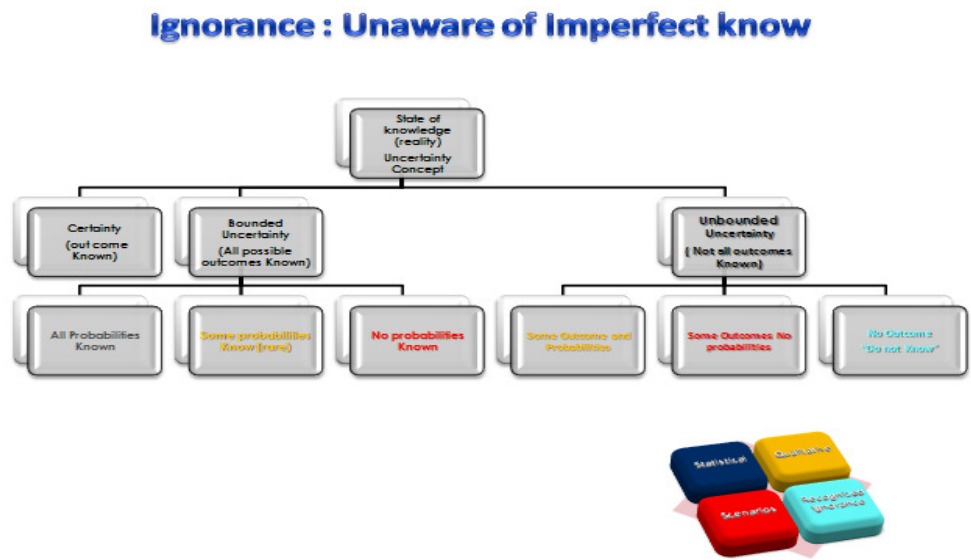

### 4.3. Nature of uncertainty

Walker et al. (2003) explain that the nature of uncertainty can be categorized into:

**Epistemic uncertainty:** the uncertainty because of imperfect knowledge and which will reduce by more study or expert advice.

**Stochastic uncertainty or ontological uncertainty:** uncertainty because of inherent variability and which is not reducible by expert advice or more study.

## 4.4. The uncertainty matrix

The uncertainty matrix can be used as a tool to get an overview of the various sources of uncertainty in a modeling study. Which will gives 'uncertainty type' in descriptions that indicate in what terms uncertainty can best be describe and the axis identifies the location or source of uncertainty and the level and nature of uncertainty.

## 6. Conclusions

**Conclusion and Feature work.**

There is not much to conclude, This is early in our study, our hope is that a systematic look towards the impact of uncertainty at module level and software development methodology, which will useful to explain, state configuration and enact software engineering process for software development process, if by that way we try to specific allocated process that will be reduce causes of software failure.

It is therefore crucial that the uncertainty is introduced in the introductory phase and tracked throughout the model study and identification, characterization of all uncertainty sources are performed jointly by the modeler, The software project manager and stakeholders in connection with the problem framing and identification of the objectives of the modeling study, which will help to developer to choose the development approaches as per level of uncertainty.

**Acknowledgements**

I would like to thanks to all staff member of computer sci. Dept. who inspire us for this work.